\documentclass[12pt]{article}
\usepackage{graphicx}
\usepackage{amsmath}
\usepackage{hyperref}

\def\ol#1{\overline{#1}}

\newcommand\pubnumber{DPF2015-89}
\newcommand\iuhetnumber{IUHET 606}
\newcommand\pubdate{\today}

\def\napoli{Physics Department\\
Indiana University, Bloomington, IN 47405, USA}

\def\Title#1{\begin{center} {\Large #1 } \end{center}}
\def\Author#1{\begin{center}{ \sc #1} \end{center}}
\def\Address#1{\begin{center}{ \it #1} \end{center}}

\newcommand\pubblock{\rightline{\begin{tabular}{l} \pubnumber\\
         \iuhetnumber\\
         \pubdate  \end{tabular}}}
\newenvironment{Abstract}{\begin{quotation}  }{\end{quotation}}
\newenvironment{Presented}{\begin{quotation} \begin{center} 
             PRESENTED AT\end{center}\bigskip 
      \begin{center}\begin{large}}{\end{large}\end{center} \end{quotation}}
\def\Acknowledgments{\bigskip  \bigskip \begin{center} \begin{large}
             \bf ACKNOWLEDGMENTS \end{large}\end{center}}

\def\beq{\begin{equation}}
\def\eeq#1{\label{#1}\end{equation}}
\def\eeqn{\end{equation}}

\def\beqa{\begin{eqnarray}}
\def\eeqa#1{\label{#1}\end{eqnarray}}
\def\eeqan{\end{eqnarray}}

\let\bar=\overbar

\def\Dslash{\not{\hbox{\kern-4pt $D$}}}
\def\dslash{\not{\hbox{\kern-2pt $\del$}}}

\def\msb{{\bar{\ssstyle M \kern -1pt S}}}

\begin{document}
\begin{titlepage}
\pubblock

\vfill
\Title{Lorentz and CPT Violation in Top-Quark Production}
\vfill
\Author{Zhi Liu}
\Address{\napoli}
\vfill
\begin{Abstract}
Signals for Lorentz and CPT violation can appear in a wide range of experiments including hadron colliders like the LHC. We present a calculation of the Lorentz-violating cross section for top-quark pair production via gluon fusion. This process dominates at the LHC, and analysis of LHC data should permit sharpening the constraints on top-quark Lorentz violation obtained recently by the D0 Collaboration. We also present a separate calculation of single-top production, which is sensitive to CPT violation. Data from the LHC can be used to measure coefficients for CPT violation in the top-quark sector for the first time.
\end{Abstract}
\vfill
\begin{Presented}
DPF 2015\\
The Meeting of the American Physical Society\\
Division of Particles and Fields\\
Ann Arbor, Michigan, August 4--8, 2015\\
\end{Presented}
\vfill
\end{titlepage}
\def\thefootnote{\fnsymbol{footnote}}
\setcounter{footnote}{0}

\section{Introduction}

Lorentz and CPT symmetries are fundamental symmetries that both the Standard Model and general relativity are based on. Up to now, a great many experiments have confirmed these symmetries. However, in the history of science, many symmetries that people had taken for granted turned out to be violated, including parity and CP. Therefore, it is possible to think about tiny violations of Lorentz and CPT symmetries. In fact, Lorentz violation may occur as a spontaneous symmetry breaking in string theories~\cite{Kostelecky:1988zi,Kostelecky:1991ak}. Thus it is becoming an important task for physicists to test these symmetries. 

To study Lorentz and CPT violation, one approach is to forget about the underlying theories that cause the violation but to come up with an effective field theory. This is more universal because it does not depend on the model we use, which is convenient for experimental searches. The framework that implement this idea is the Standard-Model Extension (SME)~\cite{Colladay:1996iz,Colladay:1998fq}, which includes all known physics and all possible Lorentz violating effects. It also has the covariance under observer Lorentz transformations, which means physics does not depend on the coordinates we choose. Under these conditions,  we can write down all possible Lorentz violating terms in the Lagrange density. Each term consists of a Lorentz violating operator contracted with a coefficient. These coefficients, which control the size of the violation, are expected to be quite small, for no deviations from Lorentz symmetry have been found. This means we can treat the coefficients perturbatively. On the other hand, in effective field theories, CPT violation leads to Lorentz violation~\cite{Greenberg:2002uu}. As a result, about half of the Lorentz violating operators in the SME also violate CPT. Therefore, the SME is also a framework to study CPT violation.

The framework of the SME allows us to predict what would happen in  experiments. In turn, experiments are able to set bounds on the coefficients. One of the most useful signals in Lorentz violation searches is the sidereal signal. As the Earth rotates on its axis, the background fields, i.e. the coefficients for Lorentz violation, which are taken to be constant in the Sun-centered frame~\cite{Bluhm:2003un,Bluhm:2001rw,Kostelecky:2002hh}, will change over time in the laboratories fixed on Earth. If the experimental systems are coupled to these fields, the results may depend on sidereal time and give sidereal signals. The sidereal day is the time it takes the Earth to complete one rotation, which is nearly 4 minutes shorter than the 24-hour mean solar day.

Various kinds of experiments have been performed in order to test Lorentz and CPT symmetries under the framework of the SME, including experiments with neutral-meson oscillations, neutrino oscillations, electroweak physics, photon, gravity, etc~\cite{table}. Many of them have achieved excellent sensitivities. In contrast, in the top-quark sector, only one measurement searching for Lorentz violation~\cite{Abazov:2012iu} and no measurement searching for CPT violation have been performed.  So this work explores the possibility of increasing the sensitivity of Lorentz violation measurement in the top quark and the prospect of testing CPT symmetry in the same context.

\section{Top pair production}

Being the heaviest quark discovered by now, the top quark plays a special role in particle physics, from the Standard Model to its extensions, including the SME~\cite{Agashe:2014kda}.  So it's of special interest to test Lorentz and CPT symmetries using the top quark~\cite{Berger:2015yha}. Up to now, the only places in the world that can produce the top quark are the Tevatron and the LHC. The top pair production, through quark fusion ($q\bar{q} \rightarrow t\bar{t}$) and gluon fusion ($gg\rightarrow t\bar{t}$), is the main production mechanism in these hadron colliders. At the Tevatron, about 85\% of the production is from quark fusion. At the LHC, gluon fusion dominates. To take advantage of the much larger cross sections of the top pair production at the LHC, it's essential to analyze how the SME modifies the cross sections of gluon fusion.

After a top-antitop pair is produced, the top and the antitop quickly decay. Therefore, the squared matrix element ($|{\cal M}|^2$) for both the production and the decay is needed for experimental analysis. Using the narrow-width approximation, $|{\cal M}|^2 $ can be written in terms of the product of three factors, which come from the production, the top decay and the antitop decay, respectively. From now on, we concentrate on the production part. The whole process is treated in Ref.~\cite{Berger:2015yha}.

The theoretical framework we use is the SME in the top-quark sector with some minor assumptions. We assume the Lorentz and CPT violation occurs only in the third-generation quarks. By field redefinitions, the dimensionless coefficient $c_{\mu\nu}$ is the only observable coefficient for the cross sections we compute. So $c_{\mu\nu}$ produces general Lorentz violation in this sense. The field redefinitions also show the antisymmetric part of $c_{\mu\nu}$ is not observable. In addition, no leading-order CPT violation can appear in top pair production~\cite{Berger:2015yha}.

The path from the theory to the experiments is like this. Starting with the SME Lagrange density, we derive the Feynman rules, which are used to calculate the squared matrix element of the production part. Combining it with the decay part gives the $|{\cal M}|^2 $ of the whole process. These results can then be used as input for experimental analysis to run simulations and to compare with experimental signals like cross sections and sidereal variations.

The D0 Collaboration has used the squared matrix element for the quark fusion to obtain the first measurement of Lorentz violation in the top-quark sector~\cite{Abazov:2012iu}. The measured $c_{\mu\nu}$ coefficient is consistent with zero with $\sim$10\% sensitivity. Given the much greater statistical power at the LHC, we expect the $c_{\mu\nu}$ coefficient can be measured to about 1\%.

Now we outline the calculations of the SME corrections to $|{\cal M}|^2$ for the gluon fusion. The Standard Model diagrams at tree level consist of three diagrams: $s$, $t$ and $u$ channels. The modified Feynman rules have insertions on vertices and propagators, which means Lorentz violating effects come in as insertions on the Standard Model diagrams. Since the Lorentz violating effects are expected to be small, we only consider the leading-order corrections. So only one insertion is needed at a time. The corrections from the vertices can be found by adding the diagrams with vertex insertions to the Standard Model diagrams, taking the modulus square and averaging/summing over spins and colors.

The calculations of the corrections from the propagators are slightly different. Instead of using insertions, we compute the full propagator. We first obtain the modified Dirac equation from the Lagrange density. The full propagator is $i$ times the factor in front of the spinor in the corresponding momentum-space equation. The modified spin sums are also needed for this calculation. They are $-i$ times the numerators of the full propagators  after we eliminate the gamma matrices in the denominators of these propagators.

The explicit expressions for the SME corrections to the squared matrix element are given in Ref.~\cite{Berger:2015yha} but are omitted here due to space limitations. When  $c_{\mu\nu}$ is factored out in the sum of the SME corrections, the remaining factor is symmetric under the interchange of $\mu$ and $\nu$, which is compatible with the antisymmetric part of $c_{\mu\nu}$ being unobservable.

\section{Single-top production}

Unlike top pair production, CPT violation is observable in single-top production, which makes it a more interesting process to consider. Since the cross sections of the single-top production are much larger at the LHC, we concentrate on the production processes at the LHC in this section.

As before, our goal is to calculate the squared matrix element for the whole process and we still use narrow-width approximation. Besides the theoretical assumption we make in Sec. 2, we further assume the only nonzero coefficient is $b_{\mu}$, which is for CPT violation. This is reasonable, because in practice it is easier to consider one coefficient at a time and set a bound on this coefficient. If we find a nonzero result, we can reconsider the interference between coefficients~\cite{Berger:2015yha}. In addition, according to Greenberg's theorem~\cite{Greenberg:2002uu}, top and antitop have the same invariant mass in our framework.

We use the same idea to calculate the SME corrections to the squared matrix element in the production part. In the Standard Model, the single-top production include the following processes: $q\ol{q'}\rightarrow t \ol{b}$ ($s$-channel); $bq \rightarrow t q'$ and $b\ol{q} \rightarrow t\ol{q'}$ ($t$-channel); $bg\rightarrow t W^-$ ($tW$ mode).  At tree level, each of the first three processes has one diagram, the $t W$ mode has two diagrams. One complication in this calculation is the modified spin sums, which can be computed using the approximate solutions to the modified Dirac equation. These solutions are given in Appendix A of Ref.~\cite{Colladay:1996iz}. To simplify the calculation, we first compute the spin sums in the zero-momentum frame of the particle, then we transform the results to the frame in which the particle has momentum $\vec{p}$. The transformation is an observer Lorentz transformation, so both the particle and the background fields $b_{\mu}$ are transformed.

The explicit expressions for the SME corrections to $|{\cal M}|^2 $ are still given in Ref.~\cite{Berger:2015yha}. One interesting result is that, if we consider single-antitop production by reversing the fermion lines in all the tree-level diagrams for single-top production, the resulting SME corrections for each process obtain an overall minus sign. This is useful, because in the Standard Model the theoretical cross sections for the $tW$ mode and for the $\ol{t}W$ mode are equal at the LHC~\cite{Bernreuther:2008ju} but the SME corrections are different. Therefore, a difference in these two cross sections is another signal for CPT violation comparing to sidereal variations.

\section{Summary}
The SME is a general framework to study Lorentz and CPT violation. Under this framework, we calculate the corrections to the squared matrix element for top pair production via gluon fusion, which allows the LHC Collaboration, including ATLAS and CMS, to increase the sensitivity of the coefficient for Lorentz violation to about 1\% comparing with the previous $\sim$10\% sensitivity. The CPT violating corrections to the squared matrix element for single-top production can be used by the LHC Collaboration to obtain the first measurement in CPT violation in the top-quark sector with an estimated sensitivity of about 5\%. The data taken with 8 TeV at the LHC can already be used to make these measurements, and using the full dataset will increase the sensitivity.

\Acknowledgments
I am grateful to Micheal S. Berger and  V. Alan Kosteleck\'y for their collaboration in this work. This work is supported partly by Department of Energy under grant number DE-SC0010120 and by the
Indiana University Center for Spacetime Symmetries
(IUCSS).

\end{document}